%% file: bdand.tex
\documentclass{caosp308}
\usepackage{graphicx}
\usepackage{natbib}
\usepackage{longtable}
\usepackage[labelsep=space,font=small,labelfont=bf]{caption}
\articleNo{291}
\pubyear{2020}
\volume{50}
\volnumber{3}
\firstpage{649}
\received{October 21, 2019}
\accepted{January 8, 2020}

\def\BibTeX{{\rm B\kern-.05em{\sc i\kern-.025em b}\kern-.08em
             T\kern-.1667em\lower.7ex\hbox{E}\kern-.125emX}}

\begin{document}

\hauthor{T.\,Pribulla et al.}

\title{Close eclipsing binary BD\,And: a triple system}

\author{
T.\,Pribulla\inst{1,2,3}\orcid{0000-0003-3599-516X}
 \and  
{\softL}.\,Hamb\'{a}lek\inst{1}
 \and 
E.\,Guenther\inst{4}
 \and 
R.\,Kom\v{z}\'{\i}k\inst{1}
 \and 
E.\,Kundra\inst{1}
 \and 
J.\,Nedoro\v{s}\v{c}\'{\i}k\inst{4}
 \and 
V.\,Perdelwitz\inst{5}
 \and 
M.\,Va\v{n}ko\inst{1}
       }

\institute{\lomnica, \email{pribulla@ta3.sk}
         \and 
           ELTE, Gothard Astrophysical Observatory, 9700 Szombathely, Szent Imre h. u. 112, Hungary
         \and
           MTA-ELTE Exoplanet Research Group, 9700 Szombathely, Szent Imre h. u. 112, Hungary
         \and
           Th\"{u}ringer Landessternwarte, Sternwarte 5, 077~78 Tautenburg, Germany 
         \and 
           Hamburger Sternwarte, Gojenbergsweg 112, 21029 Hamburg, Germany 
          }

\date{October 21, 2019}

\maketitle

\begin{abstract}
BD\,And is a fairly bright ($V$ =  10.8), active and close ($P \sim 0.9258$
days) eclipsing binary.  The cyclic variability of the apparent orbital
period as well as third light in the light curves indicate the presence of
an additional late-type component.  The principal aim is the spectroscopic
testing of the third-body hypothesis and determination of absolute stellar
parameters for both components of the eclipsing binary.  First medium and
high-resolution spectroscopy of the system was obtained.  The
broadening-function technique appropriate for heavily-broadened spectra of
close binaries was used.  The radial velocities were determined fitting the
Gaussian functions and rotational profiles to the broadening functions.  A
limited amount of photometric data has also been obtained.  Although the
photometric observations were focused on the obtaining the timing
information, a cursory light-curve analysis was also performed.  Extracted
broadening functions clearly show the presence of a third, slowly-rotating
component.  Its radial velocity is within error of the systemic velocity of
the eclipsing pair, strongly supporting the physical bond.  The observed
systemic radial-velocity and third-component changes do not support the 9
year orbit found from the timing variability.  Masses of the components of
the eclipsing pair are determined with about 0.5\% precision.  Further
characterization of the system would require long-term photometric and
spectroscopic monitoring.
\end{abstract}

\section{Introduction}
\label{intro}

Gravitational perturbations of a primordial wide binary by a third body cause so called Kozai-Lidov cycles \citep{1962AJ.....67..591K}. Cyclic variations of eccentricity of the inner binary and changes of the mutual inclination can occur if the mutual inclination, $j$, of the inner and outer orbital planes is 39\degr $< j <$ 141\degr. During the resulting close approaches of the inner binary components in the eccentric orbit tidal friction reduces orbital energy, bringing them closer. If this process is the only evolutionary channel to produce close binary stars, all close binaries must be accompanied by a third body. Unfortunately, the observational evidence is strongly biased by techniques used to detect often faint third components. While \citet{2006AJ....131.2986P}, who used several techniques and numerous observations, found at least 2/3 of contact binaries to reside in triple or multiple systems, \citet{2013ApJ...768...33R}, who searched for eclipse timing variations (light-time effect, LiTE) in the {\it Kepler} data, found only about 20\% of eclipsing binaries to be members of triple systems.

The eclipse-timing technique is sensitive to third components on long-period orbits \citep[see e.g.][]{2012AN....333..754P} but it gives only an indication because other effects can produce similar variability, e.g. mass transfer or magnetic-orbital momentum coupling \citep{1992ApJ...385..621A}. Much more conclusive is the spectroscopic \citep{2006AJ....132..769P} or visual \citep{2010AJ....139..743T} detection. There is, however, still some low probability that the additional component is physically unrelated. In the case of a wide third-body orbit, the systemic radial velocity (RV) of the binary should be close to the RV of the third component. For tighter systems mutual orbital revolution is the ultimate proof \citep{2008MNRAS.390..798P}. An analysis of eclipses in triple systems can sometimes lead not only to confirmation of multiplicity, but to accurate determination of orbital and component parameters \citep[see e.g.][]{2011Sci...331..562C}.

BD\,And (GSC 3635-1320) is a close ($P \sim 0.9258$ days), and relatively bright ($V_{\rm max}$ = 10.8) eclipsing binary. It was discovered by \citet{1938PZ......5...157} and subsequently analyzed by \citet{1938PZ......5...299}, who classified BD\,And as an Algol-type eclipsing binary with orbital period 0.462899 days and 0.46 and 0.09 mag deep minima. BD And was identified as a ROSAT source by \cite{1996ApJ...461..951S}, and it was detected by Swift \citep{2013yCat.9043....0E} and twice in the course of the XMM-Newton Slew Survey \citep{2018yCat.9053....0X}, indicating that the source has a high level of magnetic activity. Only recently, \citet{2014NewA...26...62S} obtained CCD $BVR$ photometry of the system and found that (i) the true orbital period is two times longer than determined previously, (ii) the mass ratio of the eclipsing binary is about 0.97, (iii) minima times show a cyclic variability indicating presence of a third body on about 9.6-year orbit, and (iv) light variation outside the primary eclipse may result from $\gamma$ Dor oscillations of the primary component. Subsequently \citet{2014ApJ...788..134K} analyzed extensive new $BVR$ photometry as well as published timing data. The data analysis confirmed periodic orbital period variations indicating a third component revolving on a highly eccentric orbit with $e \sim 0.76$. The light-curve (LC) variability was interpreted by dark photospheric spots on the hotter component. An LC analysis showed about 14\% third-light contribution. None of the above observations showed any flares in spite of the strong spot activity indicated by a large and variable LC asymmetry.

Unfortunately, no spectroscopic observations have been published yet, leaving the question of the system's multiplicity open. Therefore, we observed\linebreak BD\,And spectroscopically from 2016 to 2019 and obtained additional $BVI_c$ CCD photometry.

The layout of the paper is as follows. In Section~\ref{observations}, we briefly describe new spectroscopic and photometric observations. In Section~\ref{RVanalysis}, we present the analysis of the spectroscopy. The timing variability is presented in Section~\ref{lite} while LC and broadening-function (hereafter BF) modeling in Section~\ref{bflc}. Third-body parameters and orbit are discussed in Section~\ref{third}. The surface activity of the close binary in Section~\ref{activity}. The paper is concluded in Section~\ref{discussion}.

\section{New observations \label{observations} and data reduction}
\subsection{\'Echelle spectroscopy}

Medium and high-dispersion spectroscopy of BD\,And was obtained with three spectrographs. At Star\'a Lesn\'a observatory the observations were performed at the G1 pavilion with a 60cm, f/12.5 Zeiss Cassegrain telescope equipped with a fiber-fed \'echelle spectrograph eShel \citep[see][]{2011IAUS..272..282T,2015AN....336..682P}. The spectrograph has a 4150-7600 \AA~ (24 \'echelle orders), spectral range and a maximum resolving power of about $R$ = 11,000. The ThAr calibration unit provides about 100 m\,s$^{-1}$ RV accuracy. An Atik 460EX CCD camera, which has a 2749$\times$2199 array chip, 4.54 $\mu$m square pixels, read-out noise of 5.1~e$^-$ and gain 0.26e$^-$/ADU, was used as the detector. The observations were also performed with a 1.3m, f/8.36, Nasmyth-Cassegrain telescope equipped with a fiber-fed \'echelle spectrograph at Skalnat\'e Pleso. Its layout follows the MUSICOS design \citep[see][]{1992A&A...259..711B}. The spectra were recorded by an Andor iKon 936 DHZ CCD camera, with a 2048$\times$2048 array, 13.5$\mu$m square pixels, 2.7e$^-$ read-out noise and gain close to unity. The spectral range of the instrument is 4250-7375\AA~ (56 \'echelle orders) with the maximum resolution of $R$ = 38,000. Additional spectra were obtained at Th\"uringer Landessternwarte Tautenburg with the Alfred Jensch 2m telescope and coud\'e \'echelle spectrograph. These spectra cover 4510-7610 \AA~ in 51 orders. A 2.2$\arcsec$ slit was used for all observations giving $R$ = 31,500. The journal of observations is in Appendix~A. 

Because of the short orbital period of BD\,And of P$\sim$0.9258 days, the exposure times were limited to 900 seconds (about 1.1\% of the orbital period) to prevent orbital-motion smearing.

The raw data obtained with the 60cm and 1.3m telescopes were reduced using IRAF package tasks, Linux shell scripts and FORTRAN programs as described in \citet{2015AN....336..682P}. In the first step, master dark frames were produced. In the second step, the photometric calibration of the frames was done using dark and flat-field frames. Bad pixels were cleaned using a bad-pixel mask, cosmic hits were removed using the program of \citet{2004PASP..116..148P}. Order positions were defined by fitting Chebyshev polynomials to tungsten-lamp and blue LED spectra. In the following step, scattered light was modeled and subtracted. Aperture spectra were then extracted for both the object and the ThAr lamp and then the resulting 2D spectra were dispersion solved. The spectra obtained at TLS were reduced under the IRAF environment \citep[see][]{2005A&A...437..743H,2009A&A...507.1659G,2010ApJ...717..348H}.

\input{journalPH.tab}

\subsection{CCD photometry}
A limited amount of $BVI$ photometric data was obtained. Its primary goal was to better define the ephemeris for the spectroscopic observations. The data were obtained at the Star\'a Lesn\'a observatory with a 18cm f/10 auxiliary Maksutov-Cassegrain telescope attached to the Zeiss 60cm Cassegrain used to obtain the \'echelle spectroscopy (G1 pavilion). An SBIG ST10 MXE CCD camera and the Johnson-Cousins filters were used. The field of view (FoV) of the CCD camera is 28.5$\times$18.9\arcmin. Additional photometry was obtained with another 60cm Zeiss Cassegrain telescope (G2 pavilion) using a Fingerlakes ML 3041 with a back-illuminated CCD (FoV is 14.1$\times$14.1\arcmin). The filter set is also close to the Johnson-Cousins system. 

The CCD frames were photometrically reduced under the IRAF environment. First, master dark and flat-field frames were produced, then bad pixels were cleaned and the frames were photometrically calibrated. Prior to aperture photometry all frames were astrometrically solved to define the pixel to WCS\footnote{World Coordinate System} transformation. To minimize the effects of the second-order extinction, 7 nearby stars \citep[No. 9, 11, 12, 15, 16, 19, and 20 following designation in Table 1 of][]{2014ApJ...788..134K} with the total effective color $(B-V)$ = 0.586, close to that of BD\,And were chosen as comparison stars. For the largest airmass during our observations, $X$ = 1.68, the effect of the second-order extinction in the $V$ passband is $<$0.003 mag. For both telescopes the same set of comparison stars was used. A journal of the photometric observations is given in Table~\ref{journalPH_tab}.

\begin{figure}[t]
\centering
\hbox{
\includegraphics[width=6cm]{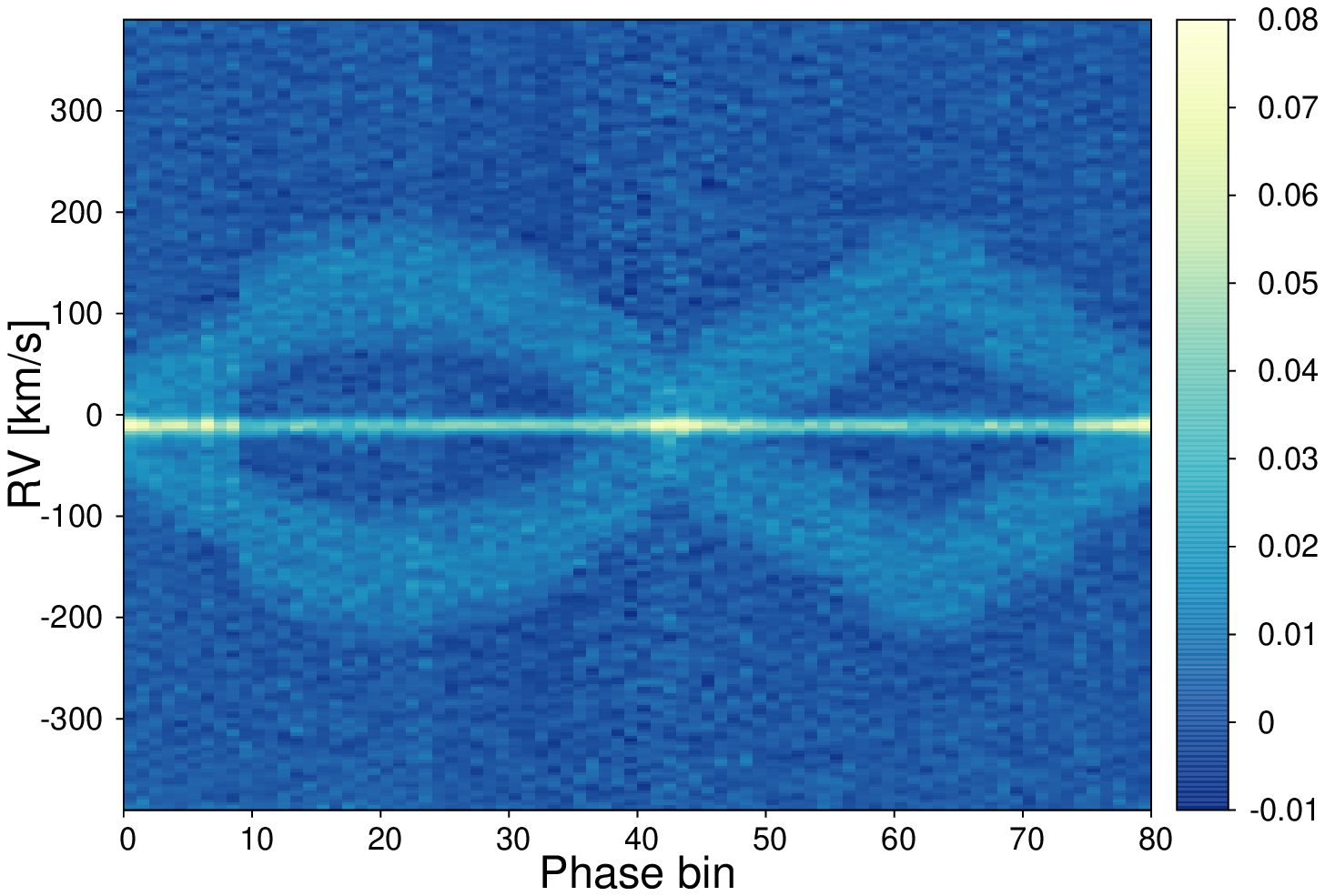}
\includegraphics[width=6cm]{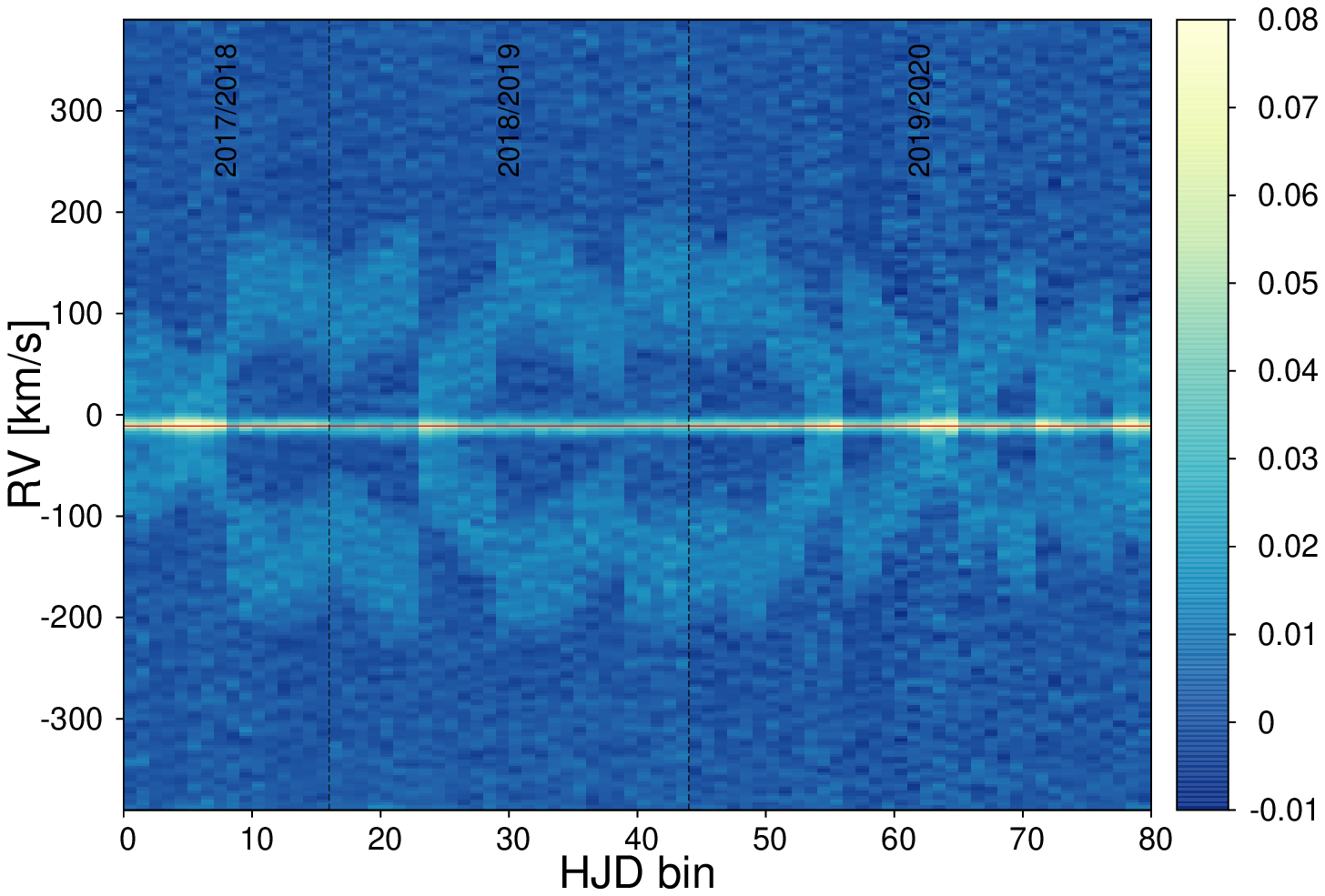}
}
\caption{Broadening functions of BD\,And extracted from the MUSICOS spectra sorted in phase (left) and ordered by date (right). The red horizontal line denotes the average radial velocity of the third component. The vertical dashed black lines separate individual observing seasons.}
\label{grayscale_fig}
\end{figure}

The heliocentric minima times determined by the \cite{1956BAN....12..327K} method are listed in Table~\ref{journalPH_tab}. The minima times are affected by the LC asymmetry, which is most marked in the $B$ passband. For example, the egress from the minimum is considerably steeper on November 7, 2017, so the minimum of light occurs slightly earlier than the spectroscopic conjunction. LC was much more symmetric in 2018.

\section{Broadening functions and radial velocities}
\label{RVanalysis}

Spectra of BD\,And were analyzed using the BF technique developed by \citet{1992AJ....104.1968R}. The BFs have been extracted in the 4900-5510 \AA~ spectral range (free of hydrogen Balmer lines and telluric lines) for all three spectrographs. The velocity step in the extracted BFs was set according to the spectral resolution. For eShel at G1 the step of $\Delta$RV = 5.8 km~s$^{-1}$ was used, for the MUSICOS and coud\'e \'echelle spectrograph at TLS, $\Delta$RV = 3.5 km~s$^{-1}$. 

BFs were extracted using HD65583 (G8V) as the template. The extracted BFs clearly show three components: two rapidly-rotating components of the eclipsing pair and a slowly rotating third component (Fig.~\ref{grayscale_fig}). The third component shows practically constant RV. 

The BFs were first modeled by triple Gaussian functions. The model Gaussian profile corresponding to the third component was then subtracted from BFs. The BFs showing non-blended components were fitted by a double rotational profile \citep[see][]{2015AN....336..682P}. The resulting RVs of the binary components are given in Appendix~A. The rotational velocities of the components measured outside eclipses are $v_1 \sin i_1$ = 69.0$\pm$0.7 km~s$^{-1}$, $v_2 \sin i_2$ = 68.8$\pm$1.2 km~s$^{-1}$ for MUSICOS at SP, and $v_1 \sin i_1$ = 70.8$\pm$1.2 km~s$^{-1}$, $v_2 \sin i_2$ = 70.7$\pm$0.8 km~s$^{-1}$ for coud\'e \'echelle at TLS. The projected rotational velocity of the third component is below the spectral resolution, thus $v_3 \sin i_3 \leq$ 8 km~s$^{-1}$. The slow rotation rate is a natural consequence of the magnetic breaking in a single late-type star.

The BFs extracted from the TLS and SP spectra are significantly less noisy and result in more precise RVs than those extracted from the G1 spectra. The RV of the third component derived from the SP and TLS spectroscopy is about $-$10.3 km~s$^{-1}$. BFs from G1 clearly show the third component, but its profile is too noisy to determine the RV reliably.

\begin{figure}[t!]
\centering
\centerline{\includegraphics[width=\columnwidth]{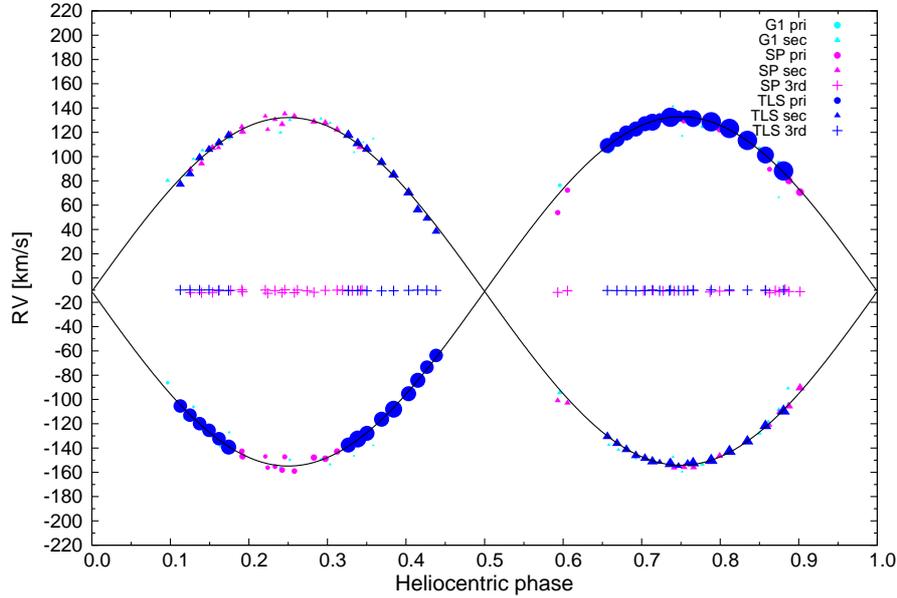}
}   
\caption{RV measurements of the primary, secondary and tertiary components of BD\,And and the corresponding Keplerian orbit fits. The point sizes scale with their weights ($1/\sigma^2$).}
\label{SPorb_fig}
\end{figure}

Our spectroscopic observations resulted in 112 RVs for all three components. Because the time range of the observations is only about 3 years, the orbital period of the eclipsing pair was adopted from equation (2) of \citet{2014ApJ...788..134K}. A circular orbit was assumed with the longitude of the periastron passage $\omega = \pi/2$. RVs of the primary and secondary component were modeled simultaneously. Errors of the RVs were estimated from equation (1) of \citet{2010ASSL..366...51H}. Assuming the same rotational velocity and spectral range, we have:
\begin{equation}
\sigma_{\rm RV} \propto \frac{1}{SNR~R^{3/2}},
\end{equation}
where $SNR$ is the signal-to-noise ratio and $R$ is the spectral resolution. Having a sufficient number of spectra for every spectrograph, the RV errors were first derived from the signal-to-noise ratios as 1/SNR. The errors were then re-scaled to give the reduced $\chi_r = 1$ for every spectrograph. 

The scaling factors for eShel at G1 are 92 and 91 for the primary and secondary component, respectively. The scaling factors for MUSICOS at SP are 37 and 36 for the primary and secondary component, respectively. The scaling factors for coud\'e \'echelle at TLS are 33 and 74 for the primary and secondary component, respectively. After re-scaling the errors, all RVs were modeled simultaneously. The resulting best parameters are given in Table~\ref{SPorb_tab} and the corresponding fits are plotted in Fig.~\ref{SPorb_fig}.

Using the inclination angle, $i = 92.72\pm 0.19$\degr, from Table 7 of \citet{2014ApJ...788..134K}, one gets $M_1$ = 1.135$\pm$0.006 M$_\odot$ and $M_2$ = 1.140$\pm$0.004 M$_\odot$. While the mass of the primary component is close to the estimate of \citet{2014ApJ...788..134K}, the secondary is about 13\% more massive and the components are of equal mass within the margins of error. Our inclination angle $i$ = 86.48(4)\degr (see Section~\ref{bflc}) gives the masses only by about 0.5\% larger than the minimum masses provided by the spectroscopic orbits.


\begin{table}[t]
\centering
\footnotesize
\caption{Spectroscopic elements of the primary and secondary components of BD\,And. The best fit is based on
112 RV measurements (30 from eShel, 50 from MUSICOS, and 32 from TLS).}
\label{SPorb_tab}
\begin{tabular}{lll}
\hline
\hline
Parameter                  & Value      & $\sigma$  \\
\hline
$P$ [d]		           & 0.92580526  &  --  \\
$T_0$ [HJD]                & 2\,457\,931.1648 &  0.0003 \\
$V_0$ [km~s$^{-1}$]        & $-$11.11    &  0.13 \\
$K_1$ [km~s$^{-1}$]        & 143.75      &  0.17 \\
$K_2$ [km~s$^{-1}$]        & 143.24      &  0.29 \\
$q = M_2/M_1$              & 1.0035      &  0.0024 \\
$a_1 \sin i$ [R$_\odot$]   & 2.629       &  0.003 \\
$a_2 \sin i$ [R$_\odot$]   & 2.620       &  0.005 \\
$M_1 \sin^3 i$ [M$_\odot$] & 1.132       &  0.005 \\
$M_2 \sin^3 i$ [M$_\odot$] & 1.136       &  0.004 \\
\hline          
\hline
\end{tabular}
\end{table}

\section{Timing variability and light-time effect}
\label{lite}

\begin{figure}[t!]
\centering
\centerline{
\includegraphics[width=\columnwidth]{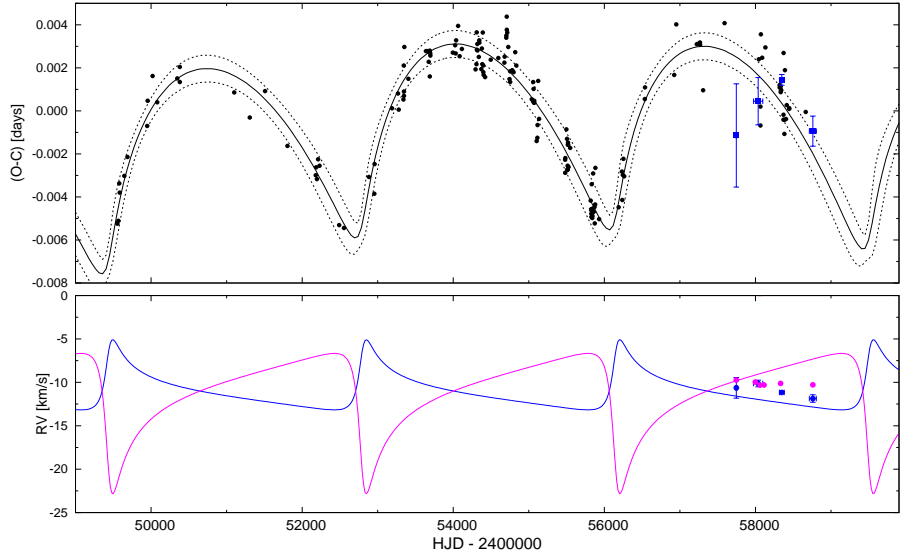}
}   
\caption{The O-C diagram for all available minima times of BD\,And (dataset a) with the best light-time effect model. The residuals are plotted for linear ephemeris $T_0$ = HJD 2\,434\,962.87143+ 0.925804848$\times E$. The blue points correspond to the mean seasonal times of the spectroscopic conjunctions. One-sigma limits are plotted with dashed lines (top). The corresponding RV-change of the eclipsing pair and the third component is also shown together with seasonal mean values (bottom). The blue line corresponds to the mass center of the eclipsing pair and the magenta line to the third component. \label{lite_fig}}
\end{figure}

BD\,And shows periodic variability of the orbital period, indicating the presence of a third component in the system. Although the cyclic variability is clearly visible (see Fig.~\ref{lite_fig}), the minima times are affected by the LC asymmetries resulting from dark photospheric spots. Moreover, the minima times were determined by different methods and obtained in different filters. Because the components are covered with cold spots (Kim et al., 2014), the spot effects are increasing in magnitude towards violet. 

The observed-computed (O-C) diagram of \citet{2014ApJ...788..134K} shows that the older photographic and visual minima timings are unusable because of the large scatter. Hence, we used CCD minima times only. In addition to the minima list of \citet{2014ApJ...788..134K} and our new observations, minima times were collected from publications of \citet{2016IBVS.6167....1P}, \citet{2015IBVS.6152....1H, 2017IBVS.6196....1H}, \citet{2015JAVSO..43...77S, 2015JAVSO..43..238S, 2018JAVSO..46...79S, 2018JAVSO..46..184S, 2019JAVSO..47..265S, 2019JAVSO..47..106S}. Because of unsure and missing errors, two datasets were analyzed: (a) CCD minima with available error estimates (149 times), and (b) all CCD minima timings assuming the same errors/weights (164 times). In the latter case the error for all minima has been set to 0.0001 days. The observed minima times were modeled assuming a LiTE and a continuous period change:

\begin{eqnarray*}
{\rm Min~I} = T_{\rm min} + P\times E + Q \times E^2 + \\ 
               + \frac{a_{12} \sin i_{12}}{c} \left[ \frac{1-e_{12}^2}{1 + e_{12} \cos \nu} \sin(\nu + \omega_{12}) + e_{12} \sin \omega_{12} \right], \\
\end{eqnarray*}

where $T_{\rm min} + P\times E + Q \times E^2$ is the quadratic ephemeris of the eclipsing pair, $a_{12} \sin i_{12}$, $e_{12}$, $\omega_{12}$, $\nu_{12}$ is the projected semi-major axis, the eccentricity, the longitude of periastron passage, and the true anomaly of the eclipsing-pair orbit around the common center of gravity, respectively. The data optimization was done assuming the linear ephemeris ($Q$ = 0, i.e., a constant period) and quadratic ephemeris (a continuous period change with $dP/dt$ = const) of the eclipsing pair.

The resulting reduced $\chi^2_r$ for a continuous-period change are 31.1 and 71.1 for the first (a) and the second dataset (b), respectively. Very high $\chi^2_r$ indicates that the published errors are underestimated or/and there is an additional intrinsic variability present in the timing data. To get reduced $\chi^2 \sim 1$ the mean observation error must be about 0.0008 days. The resulting parameters for both datasets are given in Table~\ref{lite_tab}. The corresponding fit (all CCD minima, dataset b) together with predicted and observe systemic-velocity changes is shown in Fig.~\ref{lite_fig}.

A hypothesis that the orbital period of the eclipsing pair is constant ($Q$ = 0) was tested using the F-test for case (b) where more data are available. For the false-rejection probability of $\alpha$ = 0.05 the critical value is $F$ = 3.9, while the calculated value of the $F$ statistics is 9.6. This means that the quadratic term is statistically significant. A possible cause for the observed period decrease in the eclipsing pair could be the angular momentum loss due to the magnetic breaking. A very week quadratic term was already found by \citet{2014ApJ...788..134K} from a shorter dataset (6984 vs. 9110 days). {\bf On the other hand, older photographic minima times\footnote{see e.g., http://var2.astro.cz/ocgate/} require a constant period of the eclipsing pair.}

The LC asymmetry effect on the minima times can be estimated using formula (6) of \citet{2012AN....333..754P}. Assuming an orbital period of $P$ = 0.9258 days and parameters estimated from Fig.~8 of \citet{2014ApJ...788..134K}, $D \sim 0.15P$ (minimum duration), $d \sim 0.4$ (eclipse depth), and $A_{\rm OCE} \sim 0.05$ mag we get $\Delta t$ = 0.002 days, which is comparable to the LiTE amplitude (see Fig.~\ref{lite_fig}). This means that the LiTE orbit is rather unreliable and additional information on the LC asymmetry would be needed to disentangle the spot and LiTE effects to arrive at useful orbital elements.

\begin{table}[t!]
\centering
\footnotesize
\caption{The light-time orbit of the eclipsing pair around the common center of gravity with the third component. Reduced $\chi^2_r$ is given. For dataset (b) $\sigma = 0.0008$ days was assumed for every datapoint. The time of the periastron passage, $T_{12}$, and the time of the minimum light, $T_{\rm min}$, are given without 2\,400\,000. The mass of the third component $M_3$ is given for $i_3 =  90^\circ$ (minimum mass).}
\label{lite_tab}
\begin{tabular}{lllll}
\hline
\hline
Parameter              &    Dataset (a)          &   Dataset (a)        &   Dataset (b)        &   Dataset (b)           \\
                       &    quadratic            &   linear             &   quadratic          &   linear                \\              
\hline
$T_{\rm min}$ [HJD]    &   34\,962.846(9)        &  34\,962.8696(9)     &   34\,962.851(5)     &   34\,962.8677(8)       \\   
$P$ [day]              &   0.9258070(8)          &  0.92580489(4)       &   0.9258065(5)       &   0.92580496(3)         \\  
$Q$ [day]              &   $-$4.7(17)~10$^{-11}$ &  --                  &   $-$3.7(11)~10$^{-11}$ &  --                  \\
\hline
$P_{3}$ [day]          &   3299(37)              &  3259(31)            &   3358(26)           &   3324(18)              \\  
$e_{12}$               &   0.64(9)               &  0.58(7)             &   0.74(12)           &   0.70(10)               \\
$\omega_{12}$ [deg]    &   296(7)                &  295(7)              &   308(6)             &   308(5)                \\
$T_{12}$ [HJD]         &   49\,450(80)           &  49\,510(80)         &   49\,430(60)        &   49\,490(40)           \\   
$a_{12} \sin i_{12}$ [a.u.] &  0.78(6)           &  0.74(4)             &   0.86(13)           &   0.82(9)               \\
$f(m)$ [M$_\odot$]     &   0.0057(13)            &  0.0051(9)           &   0.0077(34)         &   0.0068(21)            \\
$M_3$ [M$_\odot$]      &   0.34(3)               &  0.33(3)             &   0.38(8)            &   0.36(4)               \\
\hline
$\chi^2_r$             &   31.1                  &  32.4                &   1.111              &   1.180                 \\
d.o.f.                 &   149-8                 &  149-7               &   164-8              &   164-7                 \\
\hline
\hline
\end{tabular}
\end{table}

\section{Broadening-function and light-curve modeling}
\label{bflc}

The LC analysis of BD\,And is complicated by the surface activity resulting in wave-like distortions. A detailed analysis of $BVR$ photometry of the system was performed by \citet{2014ApJ...788..134K}. Our photometry is inferior to their data because it was focused on obtaining the timing information. Another problem is that the LC changes on the timescales as short as weeks or months. The only dataset obtained within a reasonably short time is the $BI_c$ photometry from the G2 pavilion from August 3 till September 9, 2018. Unfortunately, the observations do not cover the maximum following the primary minimum. Our best available spectroscopy (TLS data) was obtained from July 27 till August 6, 2018, thus partially overlapping with these photometric data. BFs obtained after subtraction of the third component from the triple Gaussian model were used.

The simultaneous modeling of BFs and LCs was done using the code Roche \citep[see][]{2018A&A...616A..49P}. The simultaneous analysis is crucial to lift the parameter degeneracies complicating even modeling of high-precision satellite LCs. Such is e.g. the correlation between the component radii and inclination angle \citep[see][]{2007A&A...467.1215S}. 

\begin{table}[b!]
\centering
\footnotesize
\caption{Simultaneous modeling of $BI_C$ light curves and broadening functions from the TLS spectra assuming all proximity effects using {\it Roche}. Luminosity uncertainties were computed assuming temperature errors of 100 K.}
\label{Roche_tab}
\begin{tabular}{lllll}
\hline
\hline
                           & \multicolumn{2}{c}{\it Roche} & \multicolumn{2}{c}{\citet{2014ApJ...788..134K}} \\
Parameter                  &             & $\sigma$        &             & $\sigma$  \\
\hline
$V_0$ [km~s$^{-1}$]        & $-$11.11    &  0.12           &     --      &     --    \\
$K_1 + K_2$ [km~s$^{-1}$]  & 288.04      &  0.27           &     --      &     --    \\
$q = M_2/M_1$              & 1.0017      &  0.0020         &   0.8770    &   0.0031  \\
$T_1$ [K]                  & 5550        &  --             &   5880      &     --    \\
$T_2$ [K]                  & 5522        &  3              &   5842      &   3       \\
\hline
$i$ [deg]                  & 86.48       &  0.04           &    92.72    &   0.19    \\
$R_1$ [R$_\odot$]          & 1.239       &  0.004          &    1.278    &   0.020   \\
$R_2$ [R$_\odot$]          & 1.243       &  0.005          &    1.155    &   0.018   \\
$a$ [R$_\odot$]            & 5.281       &  0.004          &    5.152    &   0.055   \\
$M_1$ [M$_\odot$]          & 1.152       &  0.003          &    1.145    &   0.053   \\
$M_2$ [M$_\odot$]          & 1.154       &  0.003          &    1.001    &   0.047   \\
$L_1$ [L$_\odot$]          & 1.31        &  0.09           &    1.75     &   0.19    \\
$L_2$ [L$_\odot$]          & 1.29        &  0.09           &    1.39     &   0.15    \\
$\log g_1$ [cgs]           & 4.313       &  0.003          &    4.284    &   0.024   \\
$\log g_2$ [cgs]           & 4.311       &  0.004          &    4.314    &   0.024   \\
\hline       
\hline   
\end{tabular}
\end{table}

Possible surface inhomogeneities were ignored due to the large phase gap in our photometric data preventing a sound analysis. The orbital period was not adjusted but fixed at $P$ = 0.92580526 days \citep[see][]{2014ApJ...788..134K}. The following parameters are used $T_0$ - time of the periastron passage, $P$ - orbital period, $i$ - inclination angle, $\Omega_1$, $\Omega_2$ - surface equipotentials, $T_1$ and $T_2$ polar temperatures of the primary and secondary component, $l_3$ - third light, $l_{12}$ - LC normalization factor, spectroscopic elements - $V_0$, $(K_1+K_2)$ and BF background and normalization factors. The limb darkening was modeled using the linear limb-darkening law and tables of \citet{1993AJ....106.2096V}. The reflection effect and gravity darkening were computed assuming convective envelopes with $\beta_1 = \beta_2$ = 0.08, and $A_1 = A_2 = 0.5$. The component fluxes were computed using model atmosphere spectral energy distribution taking into account local gravity and temperature. The polar temperature of the primary was fixed at 5550 K as found from the infrared color indices (see Section~\ref{third_sec}). Third light was not adjusted and fixed at $l_3/(l_1 + l_2) = 0.1481$ in the $B$ passband and $l_3/(l_1 + l_2) = 0.1933$ in the $I_c$ passband corresponding the the results of \citet{2014ApJ...788..134K}. A circular orbit and synchronous rotation of the components were assumed.

\begin{figure}[t!]
\centering
\includegraphics[width=12cm]{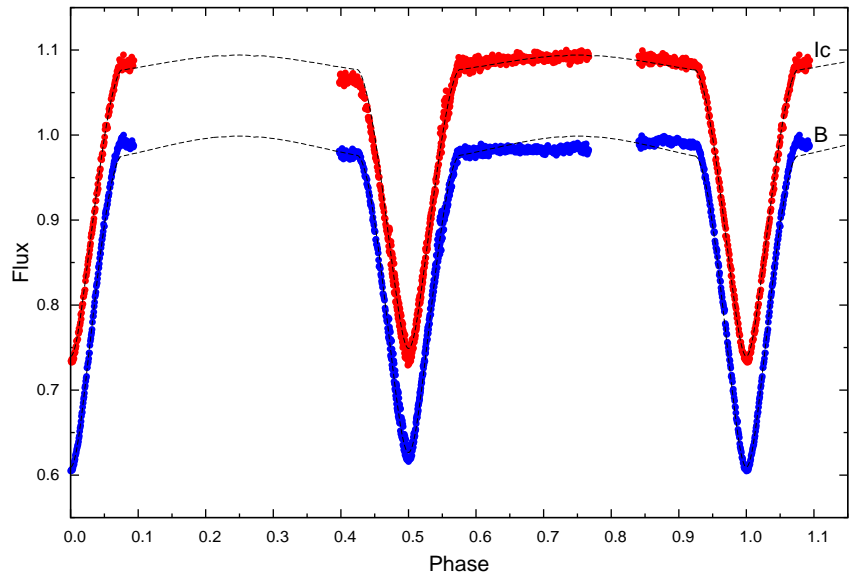}
\caption{$BI$ light curves of BD\,And obtained from August 3 till September 9, 2018 and their best fits resulting from the simultaneous modeling of light curves and broadening functions. \label{LCmodel_fig}}
\end{figure}

The $BI_c$ LCs and their best fits are shown in Fig.~\ref{LCmodel_fig}. The best parameters are listed in Table~\ref{Roche_tab}. The modeling shows that the observations (especially the $B$ data) show systematic deviations very probably caused by the surface inhomogeneities. We also note that the third light is strongly correlated to the inclination angle: the larger the third light the larger the inclination angle.

Obtaining a more reliable parameter set would require long-term monitoring of the system. Constructing the brightest LC \cite[see Section 6 of][]{2001A&A...371..997P} could provide a better reference for the modeling of individual spot-affected LCs.

\section{Third component and the outer orbit \label{third}}
\label{third_sec}

The existence of a third component on a long-period orbit was first indicated by the timing variability and third light in the LC modeling \citep[see][]{2014NewA...26...62S}. Our new spectroscopy conclusively shows the presence of an additional component.

The light contribution of the third component, $l_3/(l_1+l_2)$, determined from the Gaussian-function multi-profile fit is 0.213$\pm$0.005 (observations in an eclipsing-binary phase interval from 0.163 to 0.344). This means that the 3rd component is slightly later than the components of the eclipsing binary. The light contribution of the components in the spectrum reflects the relative strength of metallic lines (mostly Fe I, Fe II, Mg I). Because the 3rd component is of a later spectral type than the eclipsing binary and the metallic-line strength increases from G to K spectral types, the determined third light is overestimated. To correct this effect, the dependence of BF strength on $(B-V$) for the solar metallicity from Table~3 of \citet{2013AJ....146...70R} was used. Assuming that the binary components are of G1V spectral type and the third component is of G7V type \citep[][Tables 8 and 10]{2014ApJ...788..134K} and that the metallicity of all three stars is the same, the correction factor is 0.838. Thus the third-light contribution is $l_3/(l_1+l_2)$ = 0.178$\pm$0.004 or $l_3/(l_1+l_2+l_3)$ = 0.151$\pm$0.004 in the spectral range where BFs were extracted (4900-5400\AA). This is in a good agreement with the photometrically-determined value $l_3/(l_1+l_2+l_3)$ = 0.143$\pm$0.006 in the $V$ passband \citep{2014ApJ...788..134K}.

The observed combined infrared color of BD\,And is $J-K$ = 0.449$\pm$0.026. If we neglect the interstellar reddening \citep[see][]{2014ApJ...788..134K}, this corresponds to a G6-7V spectral type with one sub-type uncertainty or $T_{\rm eff}$ = 5550$\pm$100 K \citep[][]{2000asqu.book.....C}. If all three stars are main-sequence objects, the observed infrared color can be obtained having two components of the G5V spectral type and a late-type companion of the K2V spectral type. Then the combined infrared color is $J-K$ = 0.446 and third light is $l_3/(l_1+l_2)$ = 0.151, which is a bit less than our spectroscopic determination $l_3/(l_1 + l_2)$ = 0.178. Combining two main-sequence components of G6V spectral type and a K2V main-sequence star results in $l_3/(l_1 + l_2)$ = 0.171, but with a slightly redder combined color $J-K$ = 0.464. The late spectral types of the components indicated by the infrared color are, however, inconsistent with their masses, if they are main-sequence objects.

The predicted 9-year orbital revolution of the eclipsing pair around the common center of gravity is expected to cause changes of its systemic velocity. To get some constraint on the outer orbit, the RVs were divided into individual observing seasons. We kept RV semi-amplitudes, $K_1, K_2$, of the components fixed, varying only the time of the spectroscopic conjunction, $T_0$, and systemic velocity, $V_0$. Resulting systemic velocities are $V_0 = -10.63\pm 1.19$ km~s$^{-1}$ (2016/2017), $V_0 = -10.13\pm 0.38$ km~s$^{-1}$ (2017/2018), $V_0 = -11.16\pm 0.13$ km~s$^{-1}$ (2018/2019), and $V_0 = -11.86\pm 0.43$ km~s$^{-1}$ (2019/2020). The average RV of the third component per season and instrument was also computed\footnote{The only deviating point is the RV obtained from the MUSICOS data in the 2018/2019 season when better observations were obtained at TLS. This point was excluded from further considerations.}. The resulting RV data and the timing of the spectroscopic conjunction are plotted in the (O-C) diagram in Fig.~\ref{lite_fig}.

The systemic velocity and RV of the third component were observed to be identical within about 1-2 km~s$^{-1}$ and so $V_0 \sim = V_3$. This practically confirms the gravitational bond of the third component to the eclipsing pair. The observed systemic RV changes (see Section~\ref{RVanalysis}) of the eclipsing pair and the third component are, however, inconsistent with the LiTE orbit. The predicted RV difference depends on the mass ratio $M_3/(M_1 + M_2)$. Assuming that $M_3/(M_1 + M_2)$ = 0.41 \citep{2014ApJ...788..134K}, the LiTE fit predicts the RV difference to decrease from $V_0 - V_3$ = $-$1.8 km~s$^{-1}$ to $V_0 - V_3$ = $-$5.5 km~s$^{-1}$ during our spectroscopy, which is not observed. While the third component shows a constant velocity, the systemic velocity of the eclipsing pair seems to be slowly decreasing.
This means that (i) the timing variability is caused by the LiTE but the determination of $\omega_{12}$ or/and $e_{12}$ is incorrect and affected by the LC asymmetries, or (ii) the component causing the timing variability is not visible in the spectra. In the latter case, the body which causes the timing variability and eclipsing binary systemic velocity changes must be an intrinsically faint object unseen in the spectra.

A more reliable characterization of the third component would require spectroscopic observations covering the entire long-period orbit. An important requirement is the long-term RV stability of the instrument(s) at the level of at least 100-200 m~s$^{-1}$. 

\begin{figure}[t!]
\hbox{
\includegraphics[width=6cm]{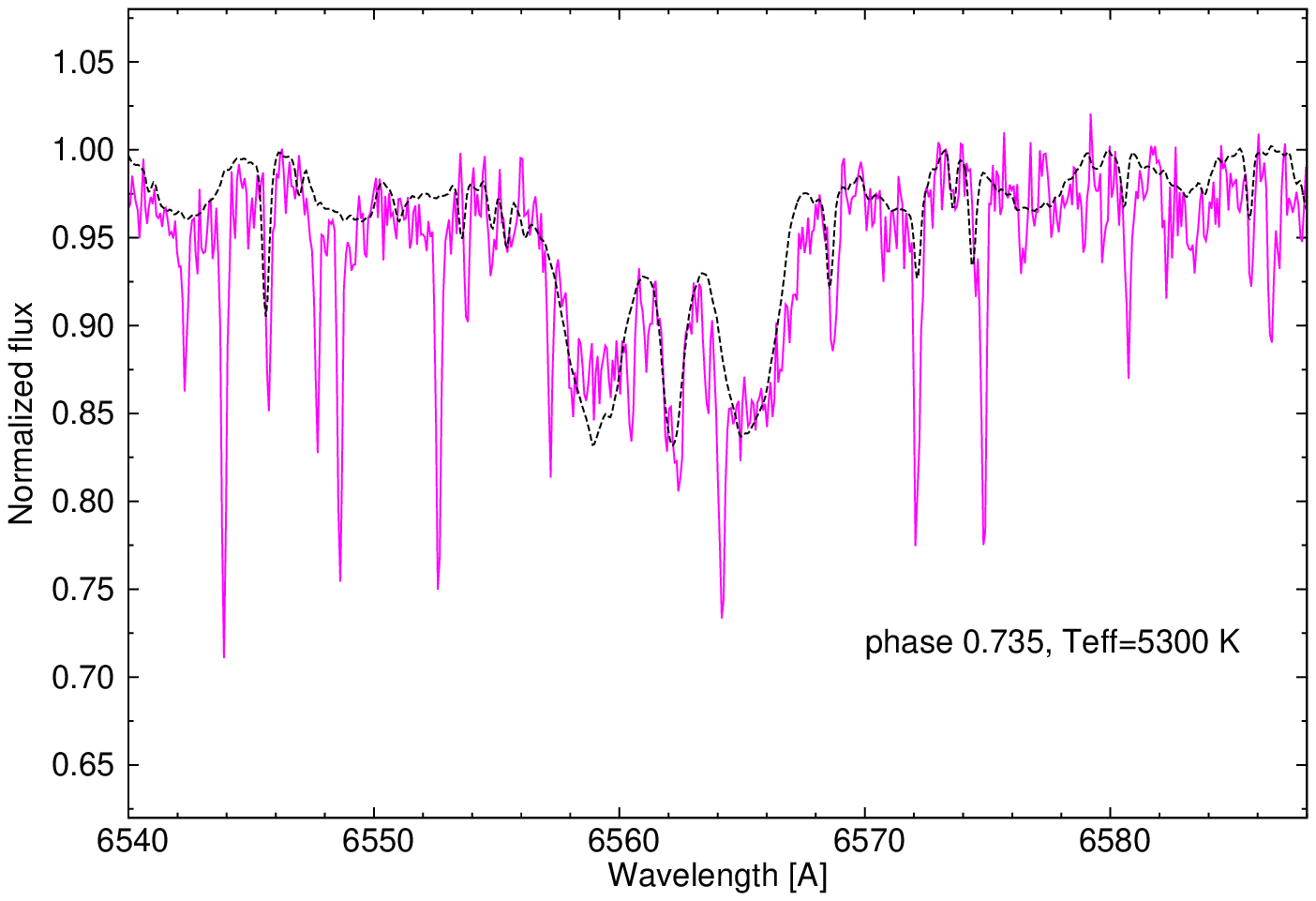}
\includegraphics[width=6cm]{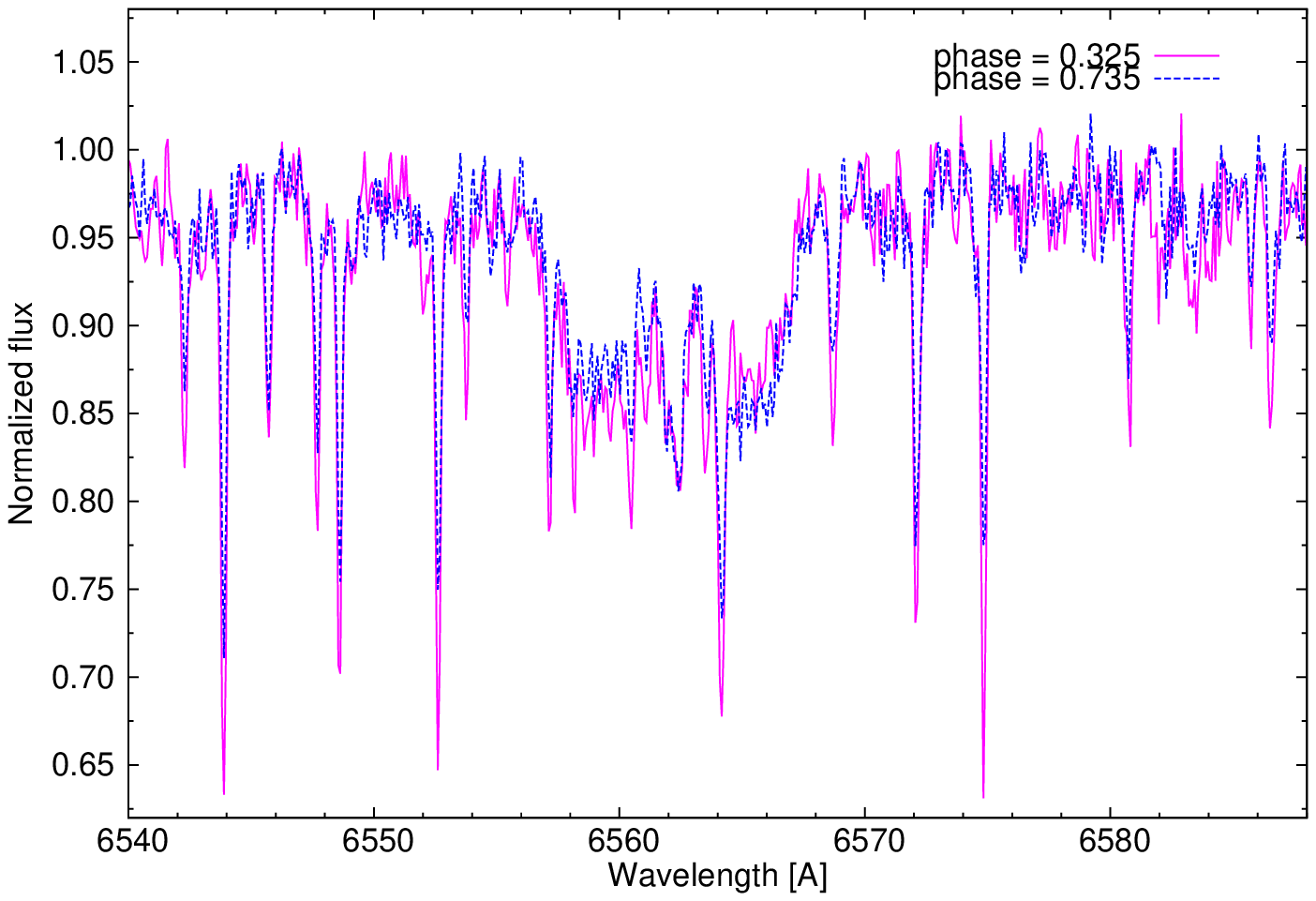}
}
\caption{Comparison of the observed H$_\alpha$ line profile (TLS spectrum from HJD~2\,458\,336.42400) and the synthetic spectrum for $T^{\rm eff}$ = 5300 K convolved with corresponding BF at phase $\phi$ = 0.735 (right). The bottom plot shows comparison of two spectra obtained at TLS at $\phi$ = 0.325 (HJD~2\,458\,330.48948) and $\phi$ = 0.735 (HJD~2\,458\,336.42400). \label{halpha_fig}}
\end{figure}

\section{Surface activity}
\label{activity}

BD\,And is composed of two solar-like components. Unlike our Sun, the components are fairly fast rotators with projected rotational velocities about 70~km\,s$^{-1}$. Thus one can expect enhanced photospheric and chromospheric activity. 

The best indicator of the surface activity are chromospheric lines Ca II H and K, $H_\alpha$ and Ca IRT \citep[see e.g.,][]{1995ApJS...96..581E,2006A&A...446.1129B,2015RAA....15..252Z}. Our spectroscopic observations cover only H$_{\alpha}$. In chromospherically active stars H$_{\alpha}$ has typically a lower equivalent width (EQW) compared to that expected for the given spectral type. To determine the extra emission, synthetic spectra have been calculated using code iSpec \citep{2014A&A...569A.111B,2014ASInC..11...85B} assuming various effective temperatures $T^{\rm eff}$ = 5100, 5300, 5500 and 5700 K, the Solar metallicity, $\log g$ = 4.3 [cgs] and the microturbulent velocity $\xi$ = 2 km~s$^{-1}$. The synthetic spectra were then convolved with extracted BFs. The resulting convolved synthetic spectra for $T^{\rm eff} \leq 5500$ K match the depth of the observed H$_{\alpha}$ line profile very well and do not indicate any extra emission (see Fig.~\ref{halpha_fig}). On the other hand, for two components of the same brightness and temperature as indicated by the combined LC and BF solution and practically equal masses (Section~\ref{RVanalysis}) we see different line depth (Fig.~\ref{halpha_fig}, bottom). The EQW is higher for the primary component. This indicates some surface activity on the secondary component lowering its EQW.

\begin{figure}[t]
\centering
\includegraphics[width=\columnwidth]{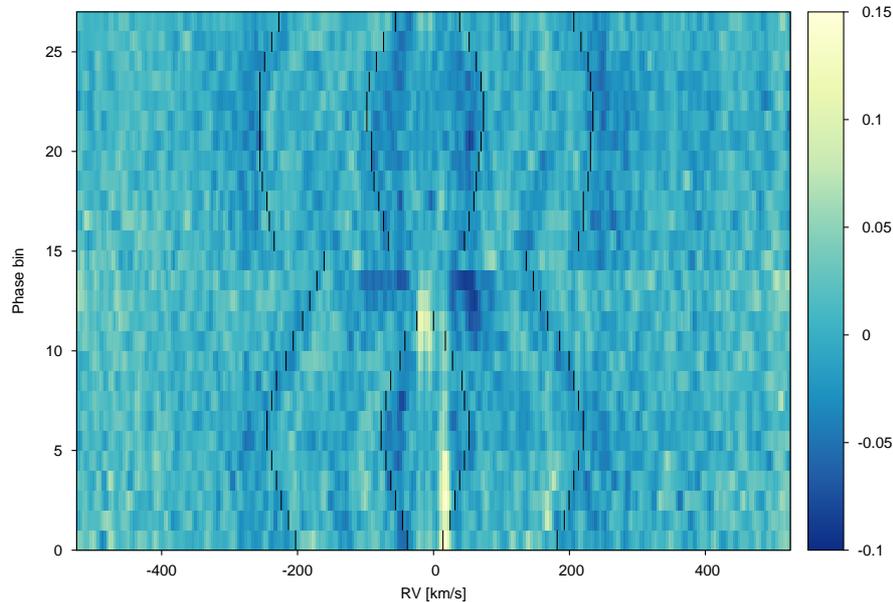}
\caption{Residuals from the BF fitting. Only TLS observations are plotted. The RV span of both components is indicated by solid black lines. The corresponding orbital-phase ranges (from bottom) are 0.111-0.173 (6 rows), 0.325-0.437 (9 rows) and 0.655-0.879 (13 rows). Only spectra from July/August 2018 are shown. \label{BFresid_fig}}
\end{figure}

We searched for direct spot signatures in the BFs extracted from the TLS high-dispersion and high SNR spectroscopy. The third component, represented by a model Gaussian function, was subtracted. The BFs were extracted from metallic lines not affected by chromospheric activity. To enhance the spots in BFs, their best fits by the Roche-code modeling (see Section~\ref{RVanalysis}) were subtracted. The resulting residuals (Fig.~\ref{BFresid_fig}) do not conclusively show the presence of spots. The analysis of BD\,And is, however, complicated by the presence the third component which cannot be completely subtracted. Moreover, the observed profiles of the components do not perfectly follow the theoretical predictions based on the solid-body rotation. 

\section{Discussion and conclusions \label{discussion}}

Extensive \'echelle spectroscopy conclusively showed that eclipsing binary BD\,And is part of a hierarchical triple system. The RVs of all three components were determined using the BF technique appropriate for heavily broadened spectra of close binaries. The power of this deconvolution method is documented by the fact that useful RVs were determined even from low SNR spectra obtained with a 60cm telescope.

The eclipsing pair is composed of two almost identical solar-type components accompanied by a late-type star. Throughout our observations the third component showed a constant RV very close to the systemic velocity of the binary, strongly supporting the gravitational bond. The orbital motion in the outer, 9-year orbit indicated by the timing variability could not be detected spectroscopically. The difference of the systemic velocity of the eclipsing pair and RV of the third component is inconsistent with the outer orbit based on the timing information. The minima times are probably shifted from spectroscopic conjunction by the surface spot activity. This negatively influences the determination of the orbital elements. Hence, future photometric observations should be obtained in the $I$ passband least affected by the spots but still accessible by silicon-based CCDs. Having both primary and secondary minima observed within a couple of days is also important. This would enable one to better quantify the spot effects on the timing information.

The identification of the third component seen in the BFs extracted from visual spectroscopy with the additional component indicated by the timing variability is spurious. The minimum mass of the third component found from the LiTE modeling ranges from 0.33 to 0.38 M$_\odot$. If it is a main-sequence star on an edge-on orbit, its light contribution would be $<$1\% \citep[see][]{2008Ap&SS.314...51X}. Then the component seen in spectra is another star on a much wider orbit.

The projected semi-major axis of the eclipsing pair around the common center of gravity found from minima timing is $a_{12} \sin i_{12}$ = 0.86$\pm$0.13 a.u. (case b with quadratic ephemeris). The corresponding systemic velocity of the eclipsing pair is predicted to range from $-$1.8 km~s$^{-1}$ to +5.5 km~s$^{-1}$. Such RV variability should easily be detected even with modest spectrographs. Long-term spectroscopic observations are crucial to determine the outer-orbit elements and mutual inclination of the inner and outer orbit. Of special importance is the determination of the mass ratio $M_3/(M_1 + M_2)$.

Interferometric observations would be hard to perform because of the low brightness of the system ($J = 9.504\pm0.022, H = 9.164\pm0.021, K = 9.055\pm 0.014$, 2MASS) even if the expected maximum separation of the eclipsing pair is about 20 mas \citep[see][]{2014ApJ...788..134K}. 

The unusable trigonometric parallax from Gaia DR2 \citep{2018arXiv180409365G}, $\pi$ = 0.13$\pm$0.63 mas is, very probably, resulting from the perturbations of an additional component (e.g., variability induced motion). It is, however, possible, that the astrometric motion will be taken into account in the final data release. In order to check the distances, we derived X-ray luminosities from the ROSAT catalog by converting count rate and hardness ratio into a physical flux with the method described by \cite{1995ApJ...450..392S}. The resulting flux, as well those from the Swift and XMM-Newton observations, were then converted into luminosities by assuming the distances in question. For a distance of 294~pc \citep{2014ApJ...788..134K}, this yields $L_X^{\rm RASS}=(7.2 \pm 1.9)\cdot 10^{30}$~erg/s, $L_X^{\rm Swift}=(2.8\pm 1.1)\cdot 10^{31}$~erg/s and $L_X^{\rm XMM}=(1.5\pm 0.5)\cdot 10^{31}$~erg/s, which corresponds to log($L_X/L_{\rm bol}$)=$-$3.3, $-$2.7 and $-$2.97, respectively. The resulting X-ray luminosities are in good agreement with each other and indicate that the system is highly active at the saturation limit of log($L_X/L_{\rm bol}$)=$-$3. This also indicates that a distance larger than $\sim 300$~pc is physically implausible, since it would implicate that all of the components of BD And are above the saturation limit.

\acknowledgements
The authors thank V.  Koll\'{a}r for his technical assistance.  This work
has been supported by the VEGA grant of the Slovak Academy of Sciences
No.\,2/0031/18, by the Slovak Research and Development Agency under the
contract No.\,APVV-015-458.  This work was also supported by the GINOP
2.3.2-15-2016-00003 of the Hungarian National Research, Development and
Innovation Office and the City of Szombathely under Agreement No. 
67.177-21/2016.  This research has made use of NASA’s Astrophysics Data
System and the SIMBAD database, operated at CDS, Strasbourg, France.  The
authors thank an anonymous referee for his/her constructive comments.

\bibliography{bdand.bib}
\clearpage
\appendix{Radial-velocity observations}
\input{journalSP.tab}

\end{document}

%% file: journalPH.tab
\begin{table*}[t!]
\footnotesize
\centering
\caption{Journal of photometric observations of BD~And. The table lists evening date (yyyymmdd), orbital phase range, No. of points, photometric filter, estimated scatter, time of the minimum light, and the instrument used. The instruments are: G1 - SBIG CCD camera in G1 pavilion at Star\'a Lesn\'a and G2 - FLI CCD camera in G2 pavilion at Star\'a Lesn\'a. The phases were computed using the same ephemeris as in Table~\ref{journalSP_tab}. \label{journalPH_tab}}
\begin{tabular}{lcrcclc}
\hline
\hline
Date     & Phases         & No.   & Filter & $\sigma$ & HJD$_{\rm min}$ & Inst. \\
         &                &       &        &          & 2\,400\,000+    &       \\
\hline
\hline
20170809 & 0.890 - 1.000  & 217   & $B$    & 0.0034   &       --        & G2 \\   
20170809 & 0.890 - 1.000  & 214   & $I$    & 0.0033   &       --        & G2 \\   
20170810 & 0.025 - 0.081  & 111   & $B$    & 0.0042   &       --        & G2 \\   
20170810 & 0.023 - 0.081  & 116   & $I$    & 0.0046   &       --        & G2 \\   
20170815 & 0.171 - 0.484  & 560   & $B$    & 0.0029   &       --        & G2 \\   
20170815 & 0.171 - 0.484  & 524   & $I$    & 0.0031   &       --        & G2 \\   
20171107 & 0.758 - 1.130  & 502   & $B$    & 0.0032   & 58065.40493(13) & G2 \\   
20171107 & 0.757 - 1.131  & 504   & $V$    & 0.0040   & 58065.40581(8)  & G2 \\   
20171115 & 0.494 - 0.604  & 134   & $V$    & 0.0038   &       --        & G1 \\   
20171116 & 0.537 - 0.548  &  14   & $V$    & 0.0049   &       --        & G1 \\   
20171124 & 0.139 - 0.214  &  88   & $V$    & 0.0045   &       --        & G1 \\   
20180720 & 0.454 - 0.580  & 208   & $B$    & 0.0036   & 58320.46604(4)  & G2 \\   
20180720 & 0.455 - 0.580  & 204   & $I$    & 0.0044   & 58320.46593(8)  & G2 \\   %
20180724 & 0.752 - 0.951  & 314   & $B$    & 0.0031   &       --        & G2 \\   
20180724 & 0.752 - 0.951  & 307   & $I$    & 0.0038   &       --        & G2 \\   
20180803 & 0.471 - 0.767  & 471   & $B$    & 0.0027   & 58334.35305(6)  & G2 \\   
20180803 & 0.471 - 0.768  & 471   & $I$    & 0.0032   & 58334.35279(16) & G2 \\   
20180804 & 0.547 - 0.621  & 117   & $B$    & 0.0030   &       --        & G2 \\   
20180804 & 0.548 - 0.620  & 116   & $I$    & 0.0037   &       --        & G2 \\   
20180807 & 0.841 - 0.094  & 394   & $B$    & 0.0029   & 58338.51898(4)  & G2 \\   
20180807 & 0.842 - 0.094  & 397   & $I$    & 0.0032   & 58338.51901(4)  & G2 \\   
20180909 & 0.000 - 0.000  & 226   & $B$    & 0.0027   & 58371.38370(6)  & G2 \\   
20180909 & 0.000 - 0.000  & 219   & $I$    & 0.0036   & 58371.38393(4)  & G2 \\   
20180921 & 0.000 - 0.000  & 317   & $B$    & 0.0025   & 58383.41891(13) & G2 \\   
20180921 & 0.000 - 0.000  & 299   & $I$    & 0.0035   & 58383.41915(10) & G2 \\   
20181123 & 0.000 - 0.000  & 350   & $B$    & 0.0026   & 58446.37438(6)  & G2 \\   
20181123 & 0.000 - 0.000  & 353   & $I$    & 0.0034   & 58446.37445(7)  & G2 \\   
\hline                                                 
\end{tabular}
\end{table*}

%% file: journalSP.tab
\footnotesize
\begin{longtable}{crrrrc}
\caption{Journal of spectroscopic observations of BD~And. The table gives heliocentric Julian date, barycentric RVs for all three components, signal-to-noise ratio and the instrument used. The instruments are: G1 - eShel spectrograph in G1 pavilion at Star\'a Lesn\'a, SP - MUSICOS spectrograph at Skalnat\'e 
Pleso, TLS - Coud\'e \'echelle at Tautenburg. Only observations where radial velocities of all three components could be determined are
listed. \label{journalSP_tab}}\\
\hline
\hline
HJD          &  RV$_1$         &  RV$_2$        & RV$_3$         &  SNR  & Inst. \\ 
$-2~400~000$ &  [km\,s$^{-1}$] & [km\,s$^{-1}$] & [km\,s$^{-1}$] &       &       \\
\hline
 57727.17144 &  103.3 & $-$137.4 &   $-$7.9 & 13 &  G1  \\
 57727.18283 &  112.5 & $-$141.4 &   $-$8.8 & 14 &  G1  \\
 57727.19328 &  115.6 & $-$140.4 &  $-$10.0 & 15 &  G1  \\
 57753.16967 &  141.4 & $-$147.4 &  $-$14.7 & 12 &  G1  \\
 57753.18048 &  117.1 & $-$159.5 &  $-$12.8 & 10 &  G1  \\
 57753.19241 &  132.9 & $-$153.9 &  $-$14.8 & 13 &  G1  \\
 57753.20425 &  124.7 & $-$153.5 &  $-$12.7 & 17 &  G1  \\
 57754.19736 &  106.8 & $-$128.0 &   $-$6.8 & 17 &  G1  \\
 57754.20782 &  102.0 & $-$118.7 &   $-$8.0 & 17 &  G1  \\
 57754.22026 &   95.3 & $-$108.4 &   $-$4.0 & 16 &  G1  \\
 57754.23120 &   90.0 &  $-$90.8 &   $-$6.5 & 14 &  G1  \\
 57966.56809 & $-$156.5 &  119.8 &  $-$14.0 & 14 &  G1  \\
 57966.57901 & $-$149.7 &  130.1 &   $-$2.9 & 13 &  G1  \\
 57968.50619 & $-$146.5 &  103.6 &   $-$4.9 & 10 &  G1  \\
 57968.51727 & $-$131.1 &  111.8 &  $-$10.7 & 10 &  G1  \\
 57968.52890 & $-$137.8 &  114.8 &   $-$8.3 & 10 &  G1  \\
 57987.51260 &  100.9 & $-$121.2 &  $-$10.8 & 11 &  G1  \\
 57987.52308 &   66.5 & $-$108.4 &  $-$19.9 & 10 &  G1  \\
 57987.53357 &   93.3 & $-$103.3 &   $-$9.0 & 10 &  G1  \\
 57991.45162 & $-$115.3 &   84.6 &  $-$13.2 & 14 &  G1  \\
 57991.46211 & $-$106.1 &   98.0 &  $-$13.6 & 13 &  G1  \\
 57991.47276 & $-$124.8 &  105.1 &  $-$10.1 & 14 &  G1  \\
 57991.48326 & $-$130.3 &  108.2 &   $-$9.4 & 14 &  G1  \\
 57991.49389 & $-$138.6 &  112.3 &   $-$8.2 & 13 &  G1  \\
 57991.50444 & $-$127.1 &  114.9 &   $-$9.0 & 12 &  G1  \\
 57992.53812 & $-$149.3 &  131.2 &  $-$12.5 & 12 &  G1  \\
 57992.54905 & $-$153.5 &  127.7 &  $-$11.8 & 12 &  G1  \\
 58028.46393 &  $-$86.2 &   80.2 &   $-$9.6 & 17 &  G1  \\
 58073.36486 &   76.4 &  $-$94.5 &   $-$8.2 & 19 &  G1  \\
 58113.26396 &  127.6 & $-$148.5 &   $-$3.4 & 11 &  G1  \\
 58060.52132 &  131.8 & $-$151.6 &  $-$10.4 & 24 &  TLS \\
 58060.53281 &  132.5 & $-$152.4 &  $-$10.4 & 29 &  TLS \\
 58060.54355 &  133.2 & $-$154.8 &  $-$10.3 & 28 &  TLS \\
 58060.55429 &  133.5 & $-$152.3 &  $-$10.2 & 28 &  TLS \\
 58081.29493 & $-$129.9 &  107.3 &  $-$10.5 & 12 &  SP  \\
 58081.30922 & $-$139.9 &  118.8 &  $-$10.1 & 12 &  SP  \\
 58081.32243 & $-$142.6 &  124.3 &   $-$9.7 & 13 &  SP  \\
 58134.17812 & $-$147.8 &  128.7 &  $-$11.7 & 15 &  SP  \\
 58134.19155 & $-$148.7 &  127.6 &  $-$10.1 & 16 &  SP  \\
 58134.20553 & $-$142.8 &  122.2 &  $-$10.0 & 15 &  SP  \\
 58134.21896 & $-$136.1 &  118.3 &  $-$10.2 & 17 &  SP  \\
 58134.23299 & $-$130.6 &  107.9 &  $-$10.3 & 17 &  SP  \\
 58327.51391 & $-$105.4 &   77.1 &   $-$9.9 & 33 &  TLS \\
 58327.52534 & $-$112.9 &   85.7 &   $-$9.8 & 35 &  TLS \\
 58327.52599 & $-$112.7 &   89.4 &  $-$12.0 & 14 &  SP  \\
 58327.53668 & $-$119.9 &   99.0 &   $-$9.9 & 34 &  TLS \\
 58327.53901 & $-$121.6 &   94.2 &  $-$12.1 & 15 &  SP  \\
 58327.54786 & $-$125.4 &  105.7 &   $-$9.8 & 33 &  TLS \\
 58327.55192 & $-$128.6 &  107.1 &  $-$11.9 & 16 &  SP  \\
 58327.55964 & $-$132.3 &  111.4 &  $-$10.0 & 33 &  TLS \\
 58327.57104 & $-$139.2 &  117.3 &  $-$10.3 & 35 &  TLS \\
 58328.51309 & $-$146.8 &  120.3 &  $-$11.3 & 15 &  SP  \\
 58328.54287 & $-$156.1 &  122.3 &  $-$12.5 & 11 &  SP  \\
 58328.55978 & $-$158.1 &  126.5 &  $-$12.0 & 14 &  SP  \\
 58328.57425 & $-$159.0 &  133.3 &  $-$12.1 & 14 &  SP  \\
 58330.48948 & $-$137.5 &  117.6 &  $-$10.4 & 38 &  TLS \\
 58330.50053 & $-$132.6 &  110.8 &  $-$10.3 & 39 &  TLS \\
 58330.51137 & $-$127.9 &  106.1 &  $-$10.5 & 38 &  TLS \\
 58330.52868 & $-$116.3 &   95.2 &  $-$10.6 & 38 &  TLS \\
 58330.54287 & $-$108.0 &   84.9 &  $-$10.5 & 39 &  TLS \\
 58331.48629 &  $-$95.3 &   70.2 &  $-$10.1 & 38 &  TLS \\
 58331.49716 &  $-$84.2 &   56.1 &  $-$10.0 & 37 &  TLS \\
 58331.50795 &  $-$73.4 &   49.0 &  $-$10.0 & 32 &  TLS \\
 58331.51879 &  $-$63.7 &   38.3 &  $-$10.2 & 32 &  TLS \\
 58334.49805 &  109.1 & $-$130.6 &  $-$10.2 & 36 &  TLS \\
 58334.50951 &  114.2 & $-$136.3 &  $-$10.3 & 36 &  TLS \\
 58334.52045 &  119.5 & $-$141.3 &  $-$10.3 & 36 &  TLS \\
 58334.53156 &  122.7 & $-$146.3 &  $-$10.6 & 36 &  TLS \\
 58334.54239 &  127.1 & $-$148.6 &  $-$10.3 & 36 &  TLS \\
 58336.40274 &  128.3 & $-$151.3 &   $-$9.9 & 39 &  TLS \\
 58336.42400 &  132.4 & $-$153.2 &   $-$9.8 & 45 &  TLS \\
 58336.45072 &  131.3 & $-$152.6 &  $-$10.0 & 44 &  TLS \\
 58336.47203 &  128.7 & $-$150.4 &   $-$9.8 & 47 &  TLS \\
 58336.49351 &  123.2 & $-$143.0 &   $-$9.9 & 46 &  TLS \\
 58336.51476 &  113.3 & $-$134.6 &  $-$10.0 & 44 &  TLS \\
 58336.53603 &  101.3 & $-$121.9 &  $-$10.0 & 43 &  TLS \\
 58336.55729 &   88.2 & $-$109.8 &  $-$10.0 & 46 &  TLS \\
 58358.51050 &   53.9 & $-$100.9 &  $-$11.8 & 13 &  SP  \\
 58358.52196 &   72.3 & $-$102.9 &  $-$10.5 & 13 &  SP  \\
 58360.49929 &  128.3 & $-$156.2 &  $-$10.6 & 14 &  SP  \\
 58360.51073 &  129.9 & $-$155.8 &  $-$10.5 & 15 &  SP  \\
 58360.52220 &  135.3 & $-$156.0 &  $-$10.4 & 15 &  SP  \\
 58360.54154 &  129.1 & $-$151.8 &  $-$11.5 & 14 &  SP  \\
 58360.55300 &  122.5 & $-$146.6 &  $-$10.8 & 16 &  SP  \\
 58360.56446 &  121.1 & $-$142.5 &  $-$10.6 & 14 &  SP  \\
 58374.49887 &   89.6 & $-$121.2 &  $-$12.2 & 12 &  SP  \\
 58374.51033 &   88.0 & $-$110.7 &  $-$11.4 & 13 &  SP  \\
 58374.52179 &   80.2 & $-$105.6 &  $-$11.1 & 18 &  SP  \\
 58374.53482 &   70.7 &  $-$90.6 &  $-$11.2 & 19 &  SP  \\
 58431.30439 & $-$146.9 &  133.3 &   $-$9.9 & 11 &  SP  \\
 58431.31585 & $-$155.3 &  130.5 &  $-$10.3 & 12 &  SP  \\
 58431.32732 & $-$147.3 &  135.2 &   $-$9.7 & 12 &  SP  \\
 58431.34228 & $-$153.1 &  127.9 &   $-$9.9 & 10 &  SP  \\
 58431.35374 & $-$150.8 &  131.5 &  $-$10.6 &  9 &  SP  \\
 58705.43381 & $-$140.9 &  117.5 &  $-$10.0 & 11 &  SP  \\
 58705.44527 & $-$137.6 &  110.3 &  $-$10.4 & 11 &  SP  \\
 58705.45672 & $-$127.6 &  106.6 &   $-$9.7 & 11 &  SP  \\
 58721.52721 &    125.2 & -149.1 &  $-$10.4 & 15 &  SP  \\
 58721.53866 &    127.7 & -151.5 &  $-$10.2 & 15 &  SP  \\
 58721.55013 &    131.1 & -153.5 &  $-$10.7 & 15 &  SP  \\
 58723.52282 &     97.3 & -120.7 &  $-$10.4 & 13 &  SP  \\
 58723.53427 &     89.8 & -113.0 &  $-$10.3 & 13 &  SP  \\
 58781.38578 &   -119.9 &  100.4 &  $-10.5$ & 17 &  SP  \\
 58781.39911 &   -116.7 &   89.1 &  $-10.5$ & 17 &  SP  \\
 58781.41057 &   -106.1 &   82.7 &  $-10.1$ & 17 &  SP  \\
 58795.31985 &    -89.0 &   67.8 &   $-9.8$ & 10 &  SP  \\
 58811.25235 &     90.3 & -110.1 &  $-10.8$ & 10 &  SP  \\
 58811.26380 &     92.8 & -122.4 &  $-10.5$ & 13 &  SP  \\
 58811.27524 &    102.8 & -122.0 &  $-10.3$ & 13 &  SP  \\
 58823.25516 &     61.4 &  -85.3 &  $-10.8$ & 15 &  SP  \\
 58823.26662 &     70.9 &  -98.0 &  $-10.2$ & 15 &  SP  \\
\hline
\hline
\end{longtable}